# Magnetic-dipolar-mode vortices and microwave subwavelength metamaterials

E.O. Kamenetskii [*], M. Sigalov, and R. Shavit

Department of Electrical and Computer Engineering,
Ben Gurion University of the Negev, Beer Sheva, Israel

**Abstract**

There has been a surge of interest in the subwavelength confinement of the electromagnetic fields. It is well known that in optics the subwavelength confinement can be obtained due to surface plasmon (quasielectrostatic) oscillations. In this paper we propose to realize the subwavelength confinement in microwaves due to dipolar-mode (quasimagnetostatic) magnon oscillations in ferrite particles. Our studies of interactions between microwave electromagnetic fields and small ferrite particles with magnetic-dipolar-mode (MDM) oscillations show strong localization of electromagnetic energy. MDM oscillations in a ferrite disk are origins of topological singularities resulting in Poynting-vector vortices and symmetry breakings of the microwave near fields. We show that new subwavelength microwave metamaterials can be realized based on a system of interacting MDM ferrite disks. The volume- and surface-wave propagation of electromagnetic signals in the proposed dense metamaterials will be characterized by topological phase variations. The MDM-particle-metamaterial concept opens a significantly new area of research. In particular, there is a perspective for creation of engineered electromagnetic fields with unique symmetry properties.

**Introduction**

The possibility of compressing electromagnetic fields in space to a degree much better than predictable by classical diffraction theory has gain widespread attention. For example, the localization of electromagnetic energy is considered as a very important phenomenon in context of applications in optical sensing and optical data communication. Use of the optical near-field characterization technique should reduce the size gap between optical and electronic devices [1, 2]. It makes possible to downscale established antenna design into the optical frequency regime [3, 4, 5]. In such optical structures, subwavelength confinement of the light takes place due to the resonant interaction between the quasielectrostatic oscillations of electrons in metal nanoparticles and planar films and the electromagnetic field.

In this paper we propose use of small ferrite particles for subwavelength confinement in microwaves. The localization of electromagnetic energy takes place due to the resonant interaction between quasi-2D ferrite disks with magnetic-dipolar [or magnetostatic (MS)] oscillations and the external electromagnetic fields. It is well known that for a given microwave frequency, the wavelength of magnetic-dipolar waves in confined magnetic structures is two-four orders of magnitude less than the wavelength of free-space electromagnetic (EM) waves [6]. As a result, one has characteristic sizes of MDM resonators much less than the free-space EM wavelength. In most cases of the present-study experiments, an excitation of MS-wave oscillations in thin-film ferrite samples is realized by microwave electric currents in microstrip transducers [7]. An interaction of the external EM fields with the MS-wave resonators is not usually a subject of these investigations. At the same time, historically, the first evidence for MS-wave oscillations was obtained by White and Solt in a microwave experiment of interaction of a small ferrite sphere with the cavity electromagnetic fields [8]. Shortly after this experiment, Dillon showed unique multiresonance MDM spectra for a ferrite disk in a microwave cavity [9]. Yukawa and Abe gave further consideration of this phenomenon in a ferrite disk [10]. For explanation of effective multiresonance interactions between the external EM fields and MS-wave oscillations, different mechanisms had been discussed [11, 12]. It was found later that a convincing model of such interactions arises from unique spectral properties of MDMs in a ferrite disk. A spectral theory of MDMs in a quasi-2D ferrite disk developed in Refs. 13 – 15, gives proper explanations of the known experimental results. One of important conclusions of the spectral theory – the fact that the MDMs in a ferrite disk can be excited not only by external microwave magnetic fields but also by external microwave electric fields – was confirmed in new microwave experiments [16 – 18].

In this paper we show that strong 3D localization of electromagnetic energy by small ferrite disks appears due to the vortex behavior of MDM oscillations. The particles with MDM oscillations are origins of the Poynting-vector vortices of the microwave near fields abutting to the disk surfaces. The Poynting-vector vortices are the regions with topological singularities. Such topological singularities are well studied in optics. For example, optical beams with phase singularities are robust structures with respect to perturbations. In such beams, one has a circulating flow of energy resulting in 2D confinement of electromagnetic energy in transversal directions [19]. In the near-field optics, due to phase singularities one obtains subwavelength transmission through narrow slits [20], novel superlenses [21], and superresolution process in metamaterials [22].

The subwavelength confinement and near-field manipulation of the electromagnetic fields is one of the main aspects attracted the concept of metamaterials [23 – 27]. We show that new subwavelength microwave metamaterials can be realized based on a system of interacting MDM ferrite disks. An array of these MDM particles with evanescent-tail chiral interactions will represent a new type of a microwave magnetic metamaterial. There are singular-microwaves metamaterials: The MDM particles are singular points with topological charges and symmetry breaking properties. As a whole, concerning the problem of microwave magnetic metamaterials, it is worth noting also that in some recent publications, interesting results on tunable photonic-band-gap structures with ferrite rods have been shown [28, 29]. These structures, however, do not have any effects of subwavelength confinement. On the other hand, in the periodic spin-wave structures called magnonic crystals, there are no effects of interactions with the external microwave EM fields [30].

**Magnetic-dipolar modes in a quasi-2D ferrite disk**

A quasi-2D MDM ferrite disk is an open resonator with a high $Q$-factor. The spectral problem for MDM oscillations is formulated for the magnetostatic-potential wave functions [13 – 15]. It was shown that the MDM disk particle can be modeled as a combined structure of a normal electric (anapole) and an in-plane rotating magnetic moments [31, 32]. As a very important property in spectral characterizations, there are vortex states of MDMs in a ferrite disk. The spectral theory gives two types of vortices. For MDMs with the anapole-moment properties, one has chiral-edge-state vortices [15], while for MDMs with the rotating-magnetic-dipole properties there are the power-flow-density vortices [32]. One of important features of the modes with the rotating-magnetic-dipole properties is a very good correspondence between analytical and numerical (based on the HFSS EM simulation program – software produced by ANSOFT Company) results of the mode characterization [31, 32]. This allows analyzing numerically the structures with such modes. For different types of the vortex states one has different mechanisms of the evanescent-tail interactions between the MDM particles. An analytical model of interacting particles was developed in Ref. [33]. In the present paper, the field structures of interacting ferrite disks are studied numerically based on the HFSS program. We analyze the EM-field confinement effects of the near-field properties of single and interacting MDM particles.

**MDM vortices in a quasi-2D ferrite disk and microwave subwavelength confinement**

To show the microwave subwavelength confinement effect, we analyze numerically the Poynting-vector distributions and the EM field structures in a rectangular waveguide with an enclosed MDM ferrite disk. Then we extend our analysis of the fields for a composition of MDM ferrite disks.

For our studies we use a ferrite disk with the following material parameters: the saturation magnetization is $4\pi M_s = 1880$ G and the linewidth is $\Delta H = 0.8$ Oe. The disk diameter is $D = 3$ mm and the disk thickness is $t = 0.05$ mm. Generally, these data correspond to the sample parameters used in microwave experiments [10, 16 – 18]. The disk is normally magnetized by a bias magnetic field $H_0 = 4900$ Oe and is placed inside a $TE_{10}$-mode rectangular waveguide symmetrically to its walls. The waveguide walls are made of copper. For a waveguide with a ferrite disk, a numerical analysis gives a multiresonance frequency characteristic of the reflection coefficient. This characteristic is represented in Fig. 1 (a). The resonance peaks are designated in succession by numbers *n*. An insertion in an upper right-hand corner of Fig. 1 (a) shows geometry of a structure with notations of power flows of incident ($\vec{P}_i$), reflected ($\vec{P}_r$), and transmitted ($\vec{P}_t$) waves. The analytically derived spectral peak positions for rotating-magnetic-dipole modes [31, 32] represented on Fig. 1 (b) are in quite good correspondence with the numerically obtained spectral peak positions in Fig. 1 (a). An insertion in the right-hand part of Fig. 1 (b) illustrates a model of a MDM ferrite disk with a rotating magnetic dipole $\vec{p}^m$ [31].

The EM-field confinement effects due to the MDM oscillations in a ferrite disk become evident from the pictures of the Poynting vector distributions. Figs. 2 (a), (b), and (c) show the Poynting vector distributions for the fields inside a waveguide on the *xz* vacuum plane situated at the distance of 150 mkm above an upper plane of a ferrite disk (in further consideration, we will conventionally call this vacuum plane as plane A). The pictures in Figs. 2 (a) and (c) correspond to the first (*f* = 8. 5225 GHz) and the second (*f* = 8. 6511 GHz) resonances, while the picture in Fig. 2 (b) is at the frequency between the resonances (*f* = 8. 5871 GHz). It is obvious that at the MDM resonant frequencies there are vortices of the Poynting vector distributions with strong subwavelength confinement of the electromagnetic energy. No such confinement is observed at non-resonance frequencies. The fact that the MDM vortices are the origins of the EM field confinement can be illustrated, additionally, by the pictures of the Poynting vector distributions inside a ferrite disk. These pictures, shown in Figs. 2 (d), (e), and (f) (which are placed in correspondence with Figs. 2 a, b, c), give evidence for the MDM power flow vortices at resonances. For non-resonance frequencies, no such MDM vortices are observed.

In Figs. 3 (a), (b), and (c) we show, respectively, the magnetic field distributions on plane A at the first-resonance frequency *f* = 8. 5225 GHz, at non-resonance frequency *f* = 8. 5871 GHz, and at the second-resonance frequency *f* = 8. 6511 GHz. To watch the dynamics, the fields are represented for two phases: $\omega t = 0°$ and $\omega t = 90°$. It is evident that at MDM resonances there are strong field concentrations and symmetry violations. No field enhancement and no symmetry violation occur at non-resonance frequencies. As a very important property of the observed pictures of the magnetic field distributions, there is an evident rotating-magnetic-dipole behavior at resonance frequencies [see Figs. 3 (a) and (c)]. Because of such rotating-magnetic-dipole behavior, predicted in Refs. [22, 23], the fields are symmetry violated and strongly confined near the disk surfaces.

**MDM vortices in a chain of quasi-2D ferrite disks and microwave subwavelength confinement**

The observed strong localization of EM energy is due to the effect when the precessing magnetic moments in a ferrite disk interact collectively, by oscillating in the MDM resonance, with the microwave fields. At the MDM resonance states, the vortex rings act as traps, providing purely subwavelength confinement of the EM fields. The interaction between the MDM ferrite disk and the electromagnetic field has two consequences. First, the presence of a circulating flow of concentrated energy presumes existence of angular EM momentum directed perpendicular to the disk plane. The second consequence is that the fields outside a ferrite disk are evanescent fields in nature: they decay exponentially with distance from the ferrite surface. All these properties offer the potential for developing new types of subwavelength microwave metamaterials. As an initial stage of an analysis of the proposed subwavelength microwave metamaterials, we consider here a chain of ferrite magnetic-dipolar-vortex particles. In such a chain, the evanescent-tail coupling between adjacent MDM resonators induces transverse dynamics and should constitute an effective waveguide structure.

In our numerical study, the chain of three quasi-2D normally magnetized ferrite disks is placed inside a $TE_{10}$-mode rectangular waveguide symmetrically to its walls. The chain is oriented along a waveguide axis. The disk diameters are 3 mm and distances between the disk axes are 3.2 mm. By virtue of quasi-magnetostatic interactions between the disks, there are splittings of MDM resonance peaks. Fig. 4 shows such splitting in the reflection coefficient characteristic for the first resonance peak [see Fig. 1 (a)]. For our analysis of a disk chain, we will choose two resonance frequencies designated in Fig. 4 as $f_1'$ and $f_1''$. In Figs. 5 (a) and (b), we show the Poynting vector distributions on plane A at resonance frequencies $f_1'$ = 8.5248 GHz and $f_1''$ = 8.5356 GHz, respectively. The pictures of the Poynting vector distributions inside every ferrite disk in Figs. 5 (c) and (d), show that the observed effect of strong subwavelength confinement of the electromagnetic fields by a disk chain is due to the MDM-resonance vortex behaviors of the particles.

For a ferrite-disk chain, the magnetic field distributions on plane A at resonance frequencies $f_1'$ = 8.5248 GHz and $f_1''$ = 8.5356 GHz and for two phases: $\omega t = 0°$ and $\omega t = 90°$ are shown in Fig. 6. There is evident field enhancement in a region near a disk chain. One can clearly observe the rotating-magnetic-dipole behavior of every ferrite disk in the chain. At two resonance frequencies, there are different relations between phases of magnetic fields in ferrite disks. At resonant frequency $f_1'$, all disks oscillate in phase [see Fig. 6 (a)]. At resonant frequency $f_1''$, there are in-phase oscillations for two extreme disks, while an interior disk oscillates in an opposite phase [see Fig. 6 (b)].

Here we considered a terminated chain of quasi-2D ferrite disks. There are standing magnetic-dipolar waves along such a chain. It is evident that for an infinite structure, an interaction between a chain of magnetic-dipole particles and the electromagnetic field will give the propagation behavior of magnetic-dipolar waves along the chain. At the same time, in the directions perpendicular to the chain, the fields will exponentially decay.

**Future directions and challenges**

   1. In optics, it was recently shown that an array of evanescent-tail-coupling nanoparticle plasmonic resonators is an effective waveguide structure with low losses [2]. This is the demonstration of non-diffraction-limited guiding of electromagnetic energy over micron- and submicron distances. Due to the heightened local fields surrounding plasmonic-resonator guiding structures, such optical devices have potential applications not only in photonics and telecommunications but also in localized biological sensing of molecules. Similarly to a system of coupled plasmonic resonators in optics, we have here effective non-diffraction-limited microwave waveguides. An array of MDM-disk resonators coupled by quasi-magnetostatic evanescent tails will be an effective microwave waveguide structure with low losses. Symmetry breakings of near fields in such structures will allow localized sensing of chiral biological and chemical objects in microwaves.
   2. Interaction of microwave fields with a MDM ferrite-disk results in Poynting-vector singularities. The power-flow-density vortices in a chain of ferrite disks [see pictures in Figs. 5 (c) and (d)] are well localized topological excitations that do not perturb the fields at large distances from the particles. Topological pointlike solutions with coreless 3D textures – the Skyrmions – are well known in nuclear and elementary particle physics [34]. Stable pointlike Skyrmions can be observed in a trapped Bose-Einstein condensate [36]. Based on an array of interacting MDM ferrite disks, one can realize subwavelength microwave photonics crystal structures with channeling of Skyrmion-like topological excitations (channeling of EM power-flow vortices).
   3. Interactions between microwave electromagnetic fields and quasi-2D ferrite disks with MDM oscillations can be limited not only to the electromagnetic wave propagation in structures with subwavelength dimensions. Dipolar-mode magnonics of ferrite particles can also help to generate and manipulate microwave electromagnetic radiation. The observed in Figs. 1 (a) and 4 strong reflections of electromagnetic power (in comparison with characteristics of an empty waveguide) at resonance frequencies of a MDM ferrite disk or a MDM ferrite-disk chain are due to certain phase relations between the phases of rotating magnetic dipoles and the phases of microwave magnetic fields in a rectangular waveguide. As a cogent argument, one can presuppose that special microwave structures with rotating magnetic fields can be realized so that the phase of a rotating magnetic dipole of a MDM ferrite disk, enclosed in such a microwave structure, will be in the opposite phase with respect to the phase of a microwave magnetic field. In this case, the rotational energy of a magnetic dipole of the ferrite-disk particle can be used for creation of an effective microwave subwavelength antenna. The microwave radiation in such a subwavelength antenna will appear as a result of collective interaction of precessing electron spins in a high-$Q$-factor MDM ferrite resonator with a magnetic field of a microwave structure.

Figure captions

**Fig. 1**. MDM resonances of a quasi-2D ferrite disk. (**a**) The numerically obtained multiresonance frequency characteristic of the reflection coefficient for a waveguide with an enclosed MDM ferrite disk. The inset shows geometry of a structure with notations of power flows of incident ($\vec{P}_i$), reflected ($\vec{P}_r$), and transmitted ($\vec{P}_t$) waves. (**b**) The analytically derived spectral peak positions for MDMs. The inset illustrates a model of a MDM ferrite disk with a rotating magnetic dipole $\vec{p}^m$.

**Fig. 2**. The field confinement originated from the MDM vortices in a ferrite disk. (**a**) The Poynting vector distributions for the fields on the *xz* vacuum plane situated at the distance of 150 mkm above an upper plane of a ferrite disk (this vacuum plane is conventionally called as plane A) at the frequency (*f* = 8. 5225 GHz) of the first resonance. (**b**) The same at the frequency (*f* = 8. 5871 GHz) between the resonances. (**c**) The same at the frequency (*f* = 8. 6511 GHz) of the second resonance. (**d**) The Poynting vector distributions inside a ferrite disk at the frequency of the first resonance. (**e**) The same at the frequency between resonances. (**f**) The same at the frequency of the second resonance.

**Fig. 3**. Evidence for correlation between strong field concentrations and symmetry violations at MDM resonances. (**a**) The magnetic field distributions on plane A at the frequency ($f$ = 8. 5225 GHz) of the first resonance for two phases: $\omega t = 0°$ and $\omega t = 90°$. (**b**) The same at the frequency ($f$ = 8. 5871 GHz) between the resonances. (**c**) The same at the frequency ($f$ = 8. 6511 GHz) of the second resonance. It is evident that at MDM resonances there are strong field concentrations and symmetry violations.

**Fig. 4.** Splittings of the MDM resonance peaks in a chain of quasi-magnetostatically interacting ferrite disks. In an analysis of a disk chain, two resonance frequencies $f_1' = 8.5248$ GHz and $f_1'' = 8.5356$ GHz are chosen.

**Fig. 5**. The field confinement originated from the MDM vortices in a chain of interacting ferrite disks. (**a**) The Poynting vector distribution on plane A at resonance frequency $f_1' = 8.5248$ GHz. (**b**) The same at resonance frequency $f_1'' = 8.5356$ GHz. (**c**) The pictures of the Poynting vector distributions inside every ferrite disk in a chain at resonance frequency $f_1' = 8.5248$ GHz. (**d**) The same at resonance frequency $f_1'' = 8.5356$ GHz.

**Fig. 6**. The magnetic field distributions vortices in a chain of interacting ferrite disks. (**a**) The magnetic field distributions on plane A at the resonance frequency $f_1' = 8.5248$ GHz for two phases: $\omega t = 0°$ and $\omega t = 90°$. (**b**) The same at the resonance frequency $f_1'' = 8.5356$ GHz. There is evident field enhancement in a region near a disk chain. One can clearly observe the rotating-magnetic-dipole behavior of every ferrite disk in the chain. At resonant frequency $f_1'$, all disks oscillate in phase. At resonant frequency $f_1''$, there are in-phase oscillations for two extreme disks, while an interior disk oscillates in an opposite phase.

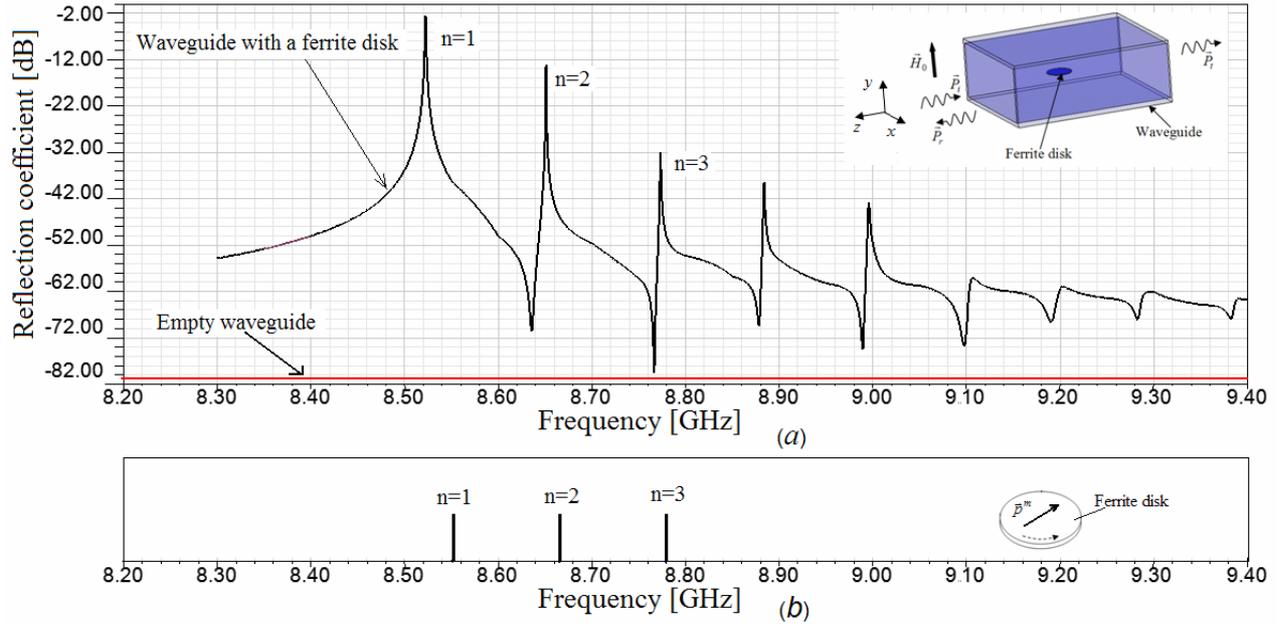

**Fig. 1**. MDM resonances of a quasi-2D ferrite disk. (**a**) The numerically obtained multiresonance frequency characteristic of the reflection coefficient for a waveguide with an enclosed MDM ferrite disk. The inset shows geometry of a structure with notations of power flows of incident ($\vec{P}_i$), reflected ($\vec{P}_r$), and transmitted ($\vec{P}_t$) waves. (**b**) The analytically derived spectral peak positions for MDMs. The inset illustrates a model of a MDM ferrite disk with a rotating magnetic dipole $\vec{p}^{\,m}$.

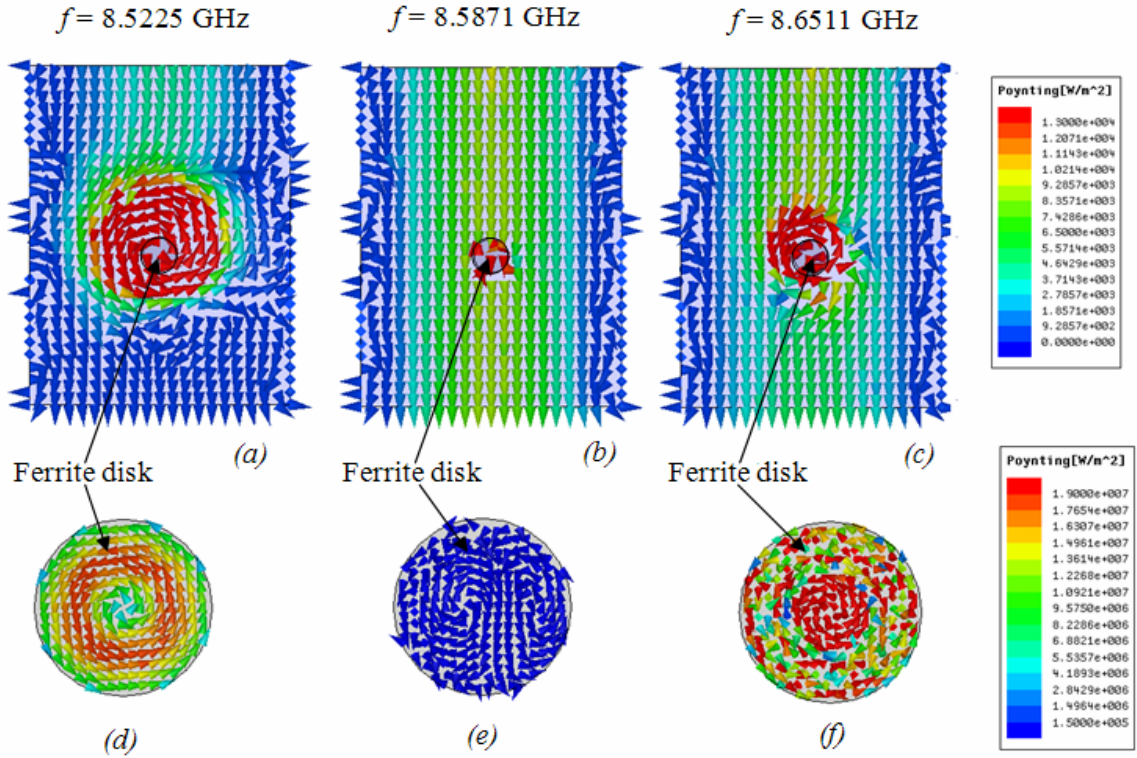

**Fig. 2**. The field confinement originated from the MDM vortices in a ferrite disk. (**a**) The Poynting vector distributions for the fields on the *xz* vacuum plane situated at the distance of 150 mkm above an upper plane of a ferrite disk (this vacuum plane is conventionally called as plane A) at the frequency (*f* = 8. 5225 GHz) of the first resonance. (**b**) The same at the frequency (*f* = 8. 5871 GHz) between the resonances. (**c**) The same at the frequency (*f* = 8. 6511 GHz) of the second resonance. (**d**) The Poynting vector distributions inside a ferrite disk at the frequency of the first resonance. (**e**) The same at the frequency between resonances. (**f**) The same at the frequency of the second resonance.

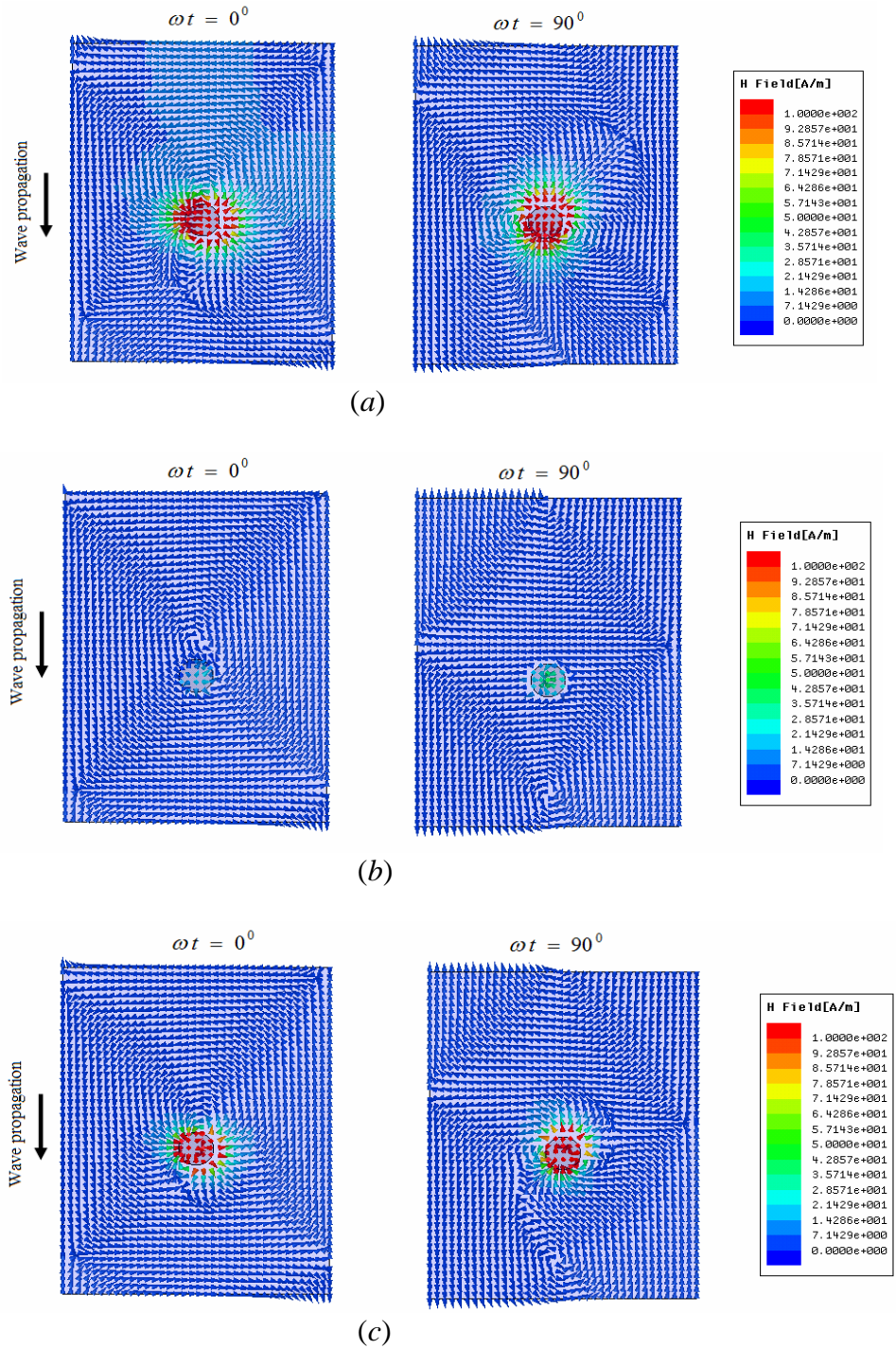

**Fig. 3**. Evidence for correlation between strong field concentrations and symmetry violations at MDM resonances. (**a**) The magnetic field distributions on plane A at the frequency ($f$ = 8. 5225 GHz) of the first resonance for two phases: $\omega t = 0°$ and $\omega t = 90°$. (**b**) The same at the frequency ($f$ = 8. 5871 GHz) between the resonances. (**c**) The same at the frequency ($f$ = 8. 6511 GHz) of the second resonance. It is evident that at MDM resonances there are strong field concentrations and symmetry violations.

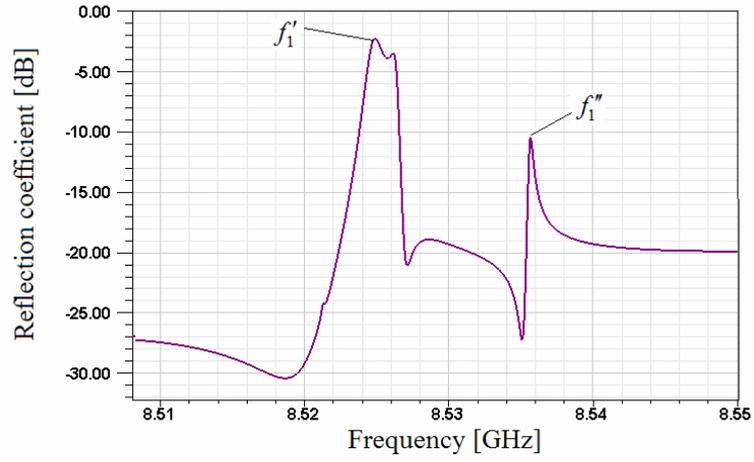

**Fig. 4.** Splittings of the MDM resonance peaks in a chain of quasi-magnetostatically interacting ferrite disks. In an analysis of a disk chain, two resonance frequencies $f_1' = 8.5248$ GHz and $f_1'' = 8.5356$ GHz are chosen.

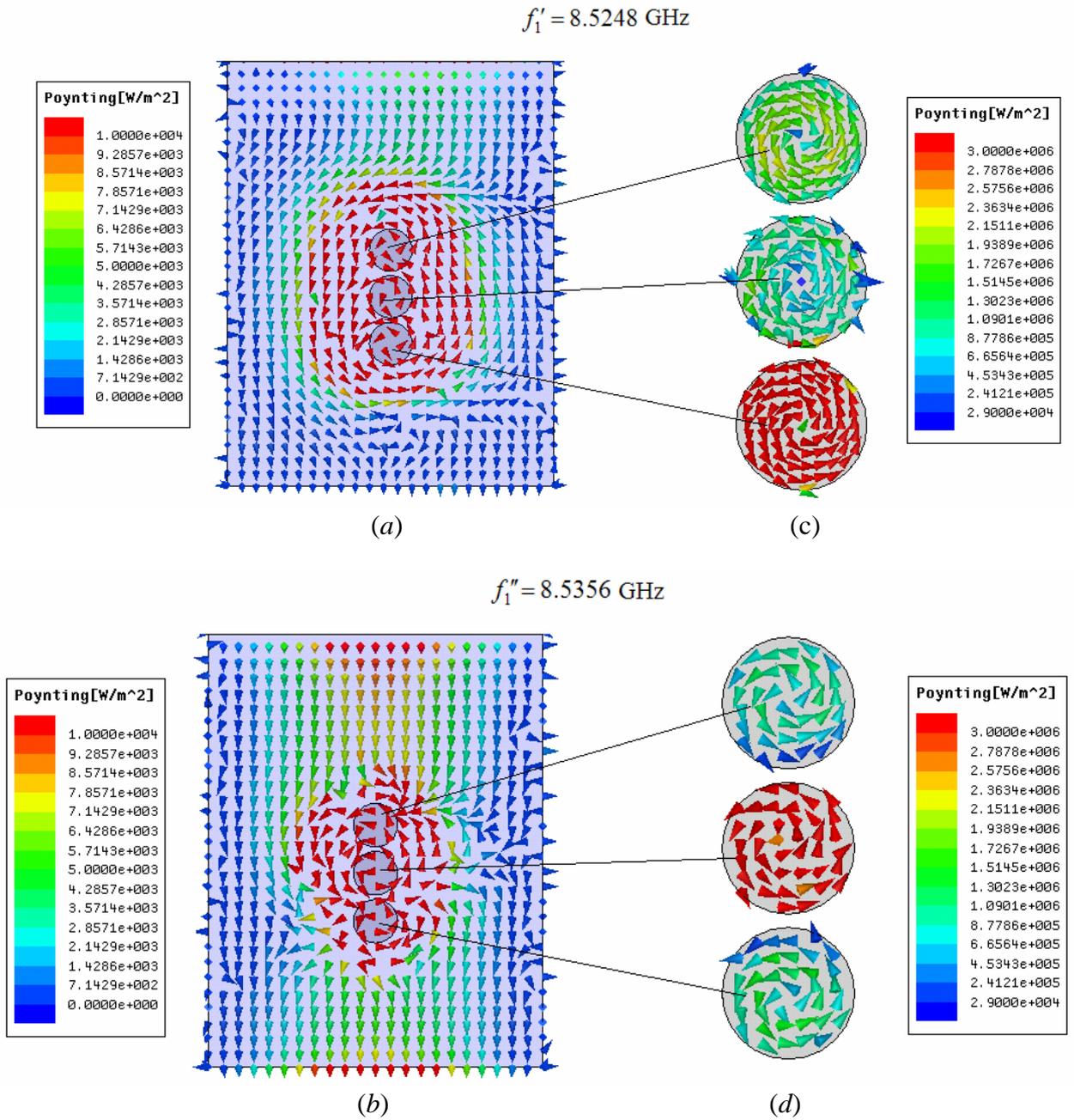

**Fig. 5**. The field confinement originated from the MDM vortices in a chain of interacting ferrite disks. (**a**) The Poynting vector distribution on plane A at resonance frequency $f_1' = 8.5248$ GHz. (**b**) The same at resonance frequency $f_1'' = 8.5356$ GHz. (**c**) The pictures of the Poynting vector distributions inside every ferrite disk in a chain at resonance frequency $f_1' = 8.5248$ GHz. (**d**) The same at resonance frequency $f_1'' = 8.5356$ GHz.

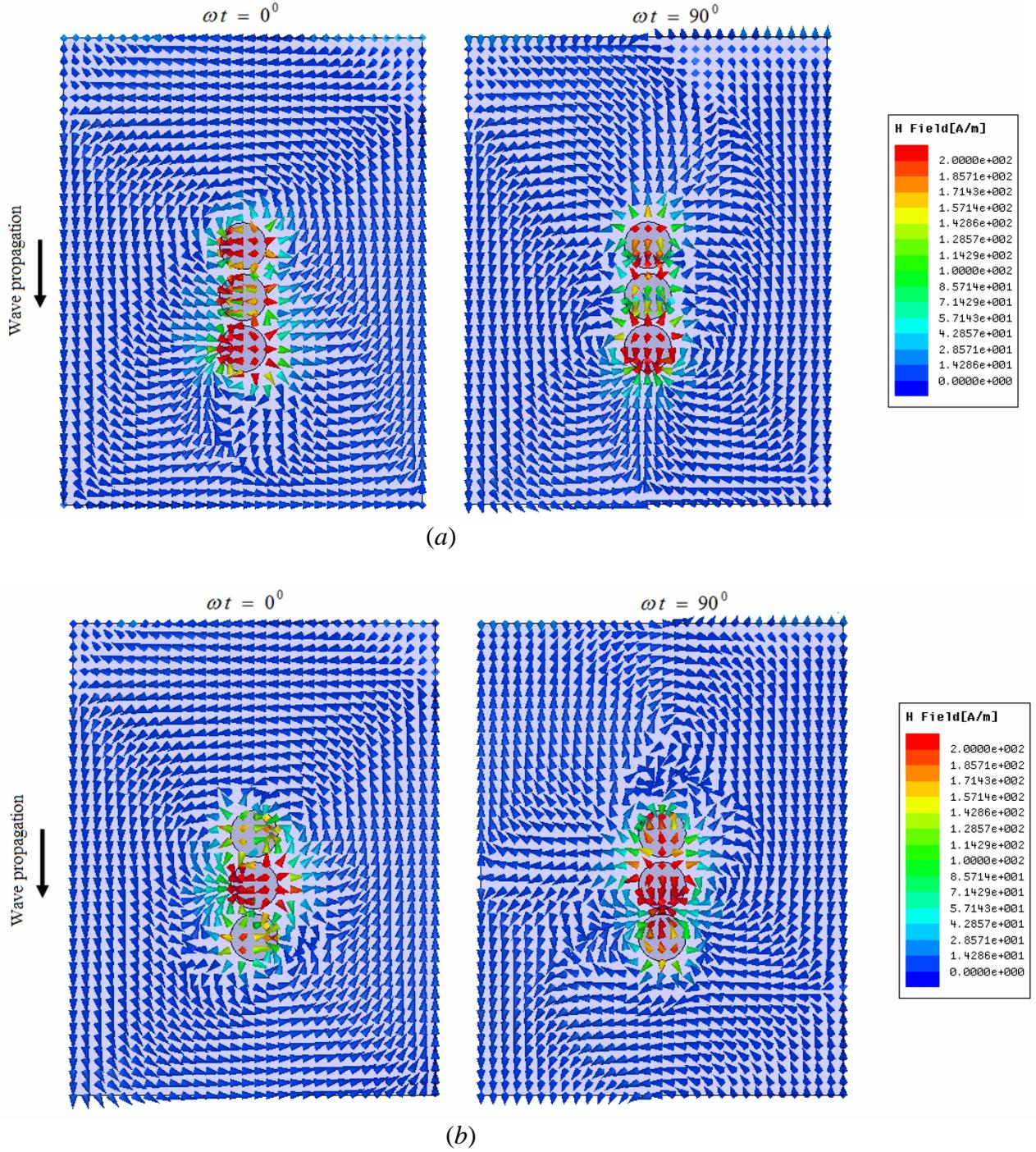

**Fig. 6**. The magnetic field distributions vortices in a chain of interacting ferrite disks. (**a**) The magnetic field distributions on plane A at the resonance frequency $f_1' = 8.5248$ GHz for two phases: $\omega t = 0°$ and $\omega t = 90°$. (**b**) The same at the resonance frequency $f_1'' = 8.5356$ GHz. There is evident field enhancement in a region near a disk chain. One can clearly observe the rotating-magnetic-dipole behavior of every ferrite disk in the chain. At resonant frequency $f_1'$, all disks oscillate in phase. At resonant frequency $f_1''$, there are in-phase oscillations for two extreme disks, while an interior disk oscillates in an opposite phase.